\documentclass[aps,prl,twocolumn,superscriptaddress]{revtex4}
\usepackage{graphicx}
\usepackage{color}

\newcommand{\be}{\begin{equation}}
\newcommand{\ee}{\end{equation}}
\newcommand{\ba}{\begin{eqnarray}}
\newcommand{\ea}{\end{eqnarray}}
\newcommand{\baa}{\begin{eqnarray*}}
\newcommand{\eaa}{\end{eqnarray*}}

\def\be{\begin{equation}}
\def\ee{\end{equation}}
\def\bea{\begin{eqnarray}}
\def\eea{\end{eqnarray}}

\def\C60{A$_x$C$_{60}$}

\def\HgCu3{HgCa$_2$Cu$_3$O$_{8+y}$}
\def\HgCu4{HgBa$_2$Ca$_3$Cu$_4$O$_{10+y}$}
\def\TlCu{Tl$_2$Ba$_2$CuO$_{6+\delta}$}
\def\TlCu3{Tl$_2$Ba$_2$Ca$_2$Cu$_3$O$_{10+y}$}
\def\TlCu4{Tl$_2$Ba$_2$Ca$_3$Cu$_4$O$_{12+y}$}

\def\BiCu3{Bi$_2$Sr$_2$Ca$_{2}$Cu$_3$O$_y$}

\def\8LSCO{La$_{1.88}$Sr$_{.12}$CuO$_4$}
\def\110LNSCO{La$_{1.5}$Nd$_{0.4}$Sr$_{0.1}$CuO$_{4}$}
\def\stage4LCO{La$_{2}$CuO$_{4+\delta}$}
\def\Y248{YBa$_2$Cu$_4$O$_8$}

\def\NbSe2{NbSe$_2$}
\def\TaSe2{TaSe$_2$}
\def\TiSe2{TiSe$_2$}

\begin{document}
\title{Electronic Physics and Possible Superconductivity in Layered Orthorhombic Cobalt Oxychalcogenides }

\author{Congcong Le}
\affiliation{Beijing National Laboratory for Condensed Matter Physics, and Institute of Physics,
Chinese Academy of Sciences, Beijing 100190, China}
\author{Shengshan Qin}
\affiliation{Beijing National Laboratory for Condensed Matter Physics, and Institute of Physics,
Chinese Academy of Sciences, Beijing 100190, China}
\author{Jiangping Hu}\email{jphu@iphy.ac.cn }
 \affiliation{Beijing National Laboratory for Condensed Matter Physics, and Institute of Physics, Chinese Academy of Sciences, Beijing 100190, China}
\affiliation{Collaborative Innovation Center of Quantum Matter, Beijing, China}
\affiliation{Kavli Institute of Theoretical Sciences, University of Chinese Academy of Sciences,  Beijing 100049, China}

\begin{abstract}
We suggest that Cobalt-Oxychalcogenide layers constructed by vertex sharing CoA$_2$O$_2$ (A=S,Se,Te) tetrahedra, such as BaCoAO, are strongly correlated multi-orbital electron systems that can provide important clues on the cause of unconventional superconductivity. Differing from cuprates and iron-based superconductors, these systems lack of the C$_4$ rotational symmetry. However,  their parental compounds  possess antiferromagnetic(AFM) Mott insulating  states through pure   superexchange interactions and  the low energy physics near Fermi surfaces  upon doping is also attributed mainly to the three t$_{2g}$ orbitals that dominate  the AFM interactions. We derive a low energy effective model for these systems and  predict that a d-wave-like superconducting state with reasonable high transition temperature can emerge by suppressing  the AFM ordering even if the pairing symmetry can not be classified by the rotational symmetry any more.
\end{abstract}

\pacs{75.85.+t, 75.10.Hk, 71.70.Ej, 71.15.Mb}

\maketitle

%
%


 In the past three decades, two families of unconventional high temperature superconductors (high -T$_c$), cuprates\cite{Bednorz1986} and iron-based superconductors\cite{Kamihara2008-jacs}, were discovered.  Enormous research efforts have been devoted to understand their origin of high T$_c$.  However, lacking of general principles that can guide us to search for new  high T$_c$ materials is still the major barrier in reaching a consensus in this field.

  Recently, we have shown that  the two known families of  high-T$_c$  share a common electronic property--those d-orbitals that make the strongest in-plane d-p couplings   are isolated near Fermi surface energy\cite{Hu-tbp,Hu-genes, Hu-s-wave}. This special electronic character separates the two families of high $T_c$ materials from other correlated electronic materials. We have further argued that this character can only be realized in very limited special cases so that it can serve as the clue to guide us in searching for next families of unconventional high T$_c$ materials. Following this guide, we have found that the condition can be realized in two special electronic structures. The first one is  a two dimensional hexagonal lattice formed by edge-shared trigonal biprymidal complexes\cite{Hu-tbp} and the second one is a two dimensional square lattice formed by vertex-shared tetrahedra complexes\cite{Hu2016}. In both cases,  the condition is satisfied when the electron filling configuration in cation atoms is near d$^7$, which suggests that  layered Co$^{2+}$/Ni$^{3+}$ based materials  can be promising high $T_c$ families.  Furthermore, following the Hu-Ding principle\cite{Huding2012} of the pairing symmetry selection,  the $d+id$ and  $d$ pairing symmetries are  selected respectively in these two families. Confirming the above predictions can promisingly settle the elusive unconventional high T$_c$ mechanism and open the window  for theoretically designing high T$_c$ materials. However, materials with above characteristics are new materials and it is  unknown whether they can be synthesized successfully.

  \begin{figure}
\centerline{\includegraphics[width=0.5\textwidth]{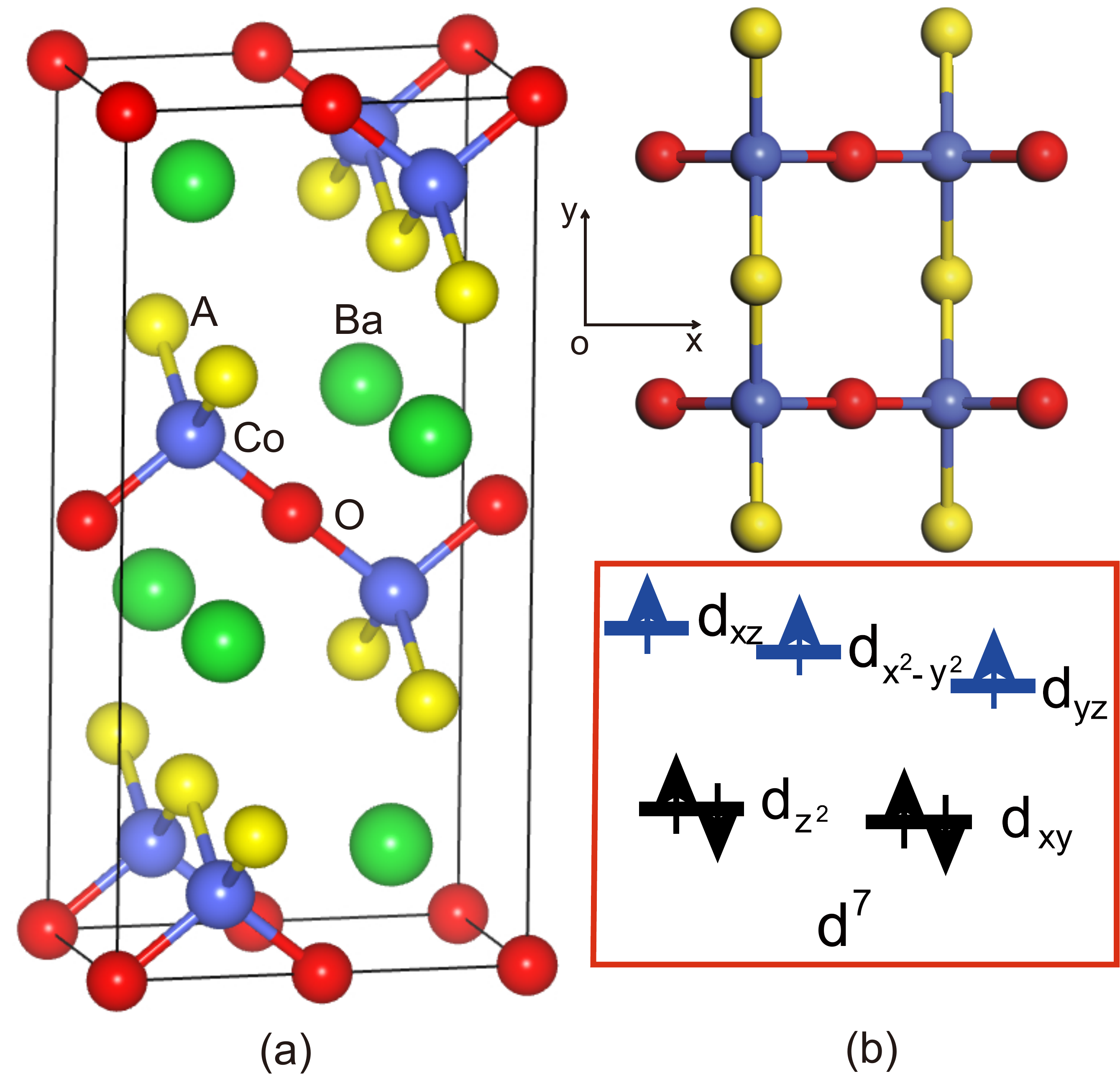}}
\caption{(color online) (a) The Crystal structure of BaCoAO (A=S, Se). (b) The Crystal field splitting of BaCoAO.
\label{lattice} }
\end{figure}
 In this paper, we turn our attention to  a family of   materials which are electronically close to our second proposed structures\cite{Salter2016,Valldor2015}. In our second proposal,  the electronic structure stems from a two dimensional layer with vertex sharing tetrahedra, such as Co(S,Se)$_4$\cite{Hu2016}. Such a square lattice structure has not been founded in current material database.  However, we notice that  there are orthorhombic layered materials that have been synthesized recently, such as BaCoSO, in which each CoSO layer  is constructed by vertex sharing mixed-anion tetrahedron complexes, CoA$_2$O$_2$ (A=S,Se), as shown in Fig.\ref{lattice}(a).  In this layer structure,   the Co chains along b axis are formed through S/Se atoms and  the staggered Co chains along a-axis are formed through O atoms. In a CoA$_2$O$_2$ tetrahedron, the Co$^{2+} $ ion is not  located in a perfect tetrahedron environment and  the formed layer lacks of C$_4$ rotational symmetry. Compared with  the second proposed structures\cite{Hu2016},  the three t$_{2g}$ orbitals here can be highly non-degenerate, the electronic structures are intrinsically nematic, the hybridization between t$_{2g}$ and e$_{g}$ orbitals  is  slightly increased, and so does the hybridization between  the  t$_{2g}$ orbitals. Thus, the electronic physics in these systems can be much richer and more complex.  The complexity  makes  theoretical studies difficult to predict possible emergence of superconductivity in these materials. However, if the major electronic physics is not altered and  the electronic conditions can be optimized,  superconductivity with reasonable high $T_c$  should be developed by suppressing AFM ordering.

In this Letter, we combine density functional theory (DFT), numerical and analytic methods to investigate the electronic physics of these materials.  We found that the  e$_g$ orbitals and the t$_{2g}$ orbitals in these materials are  well separated by the crystal field as shown in Fig.\ref{lattice}(b) so that the t$_{2g}$ orbitals    still dominate the low energy physics near Fermi energy.  A low energy effective model is derived for the  two dimensional CoAO layer. Without doping, the system is in a high spin AFM Mott insulating state.  By introducing charge carriers,  the three t$_{2g}$ orbitals make dominant contribution near Fermi energy in a paramagnetic metallic state. Among the three t$_{2g}$ orbitals, the d$_{yz}$ orbital forms a quasi-one dimensional hole-like Fermi surface in a folded-out Brillouin zone (BZ) with respect to the one Co unit cell and is relatively decoupled from the other two t$_{2g}$ orbitals. The other two  t$_{2g}$  orbitals  are strongly coupled with each other along the a (x) axis and depending on doping, they  can form an electron pocket at the zone corner and  a hole pocket at the zone center.  If  the AFM ordering is suppressed,  we predict that a d-wave like superconducting state with reasonable high transition temperature can be developed.

\begin{figure}
\centerline{\includegraphics[width=0.5\textwidth]{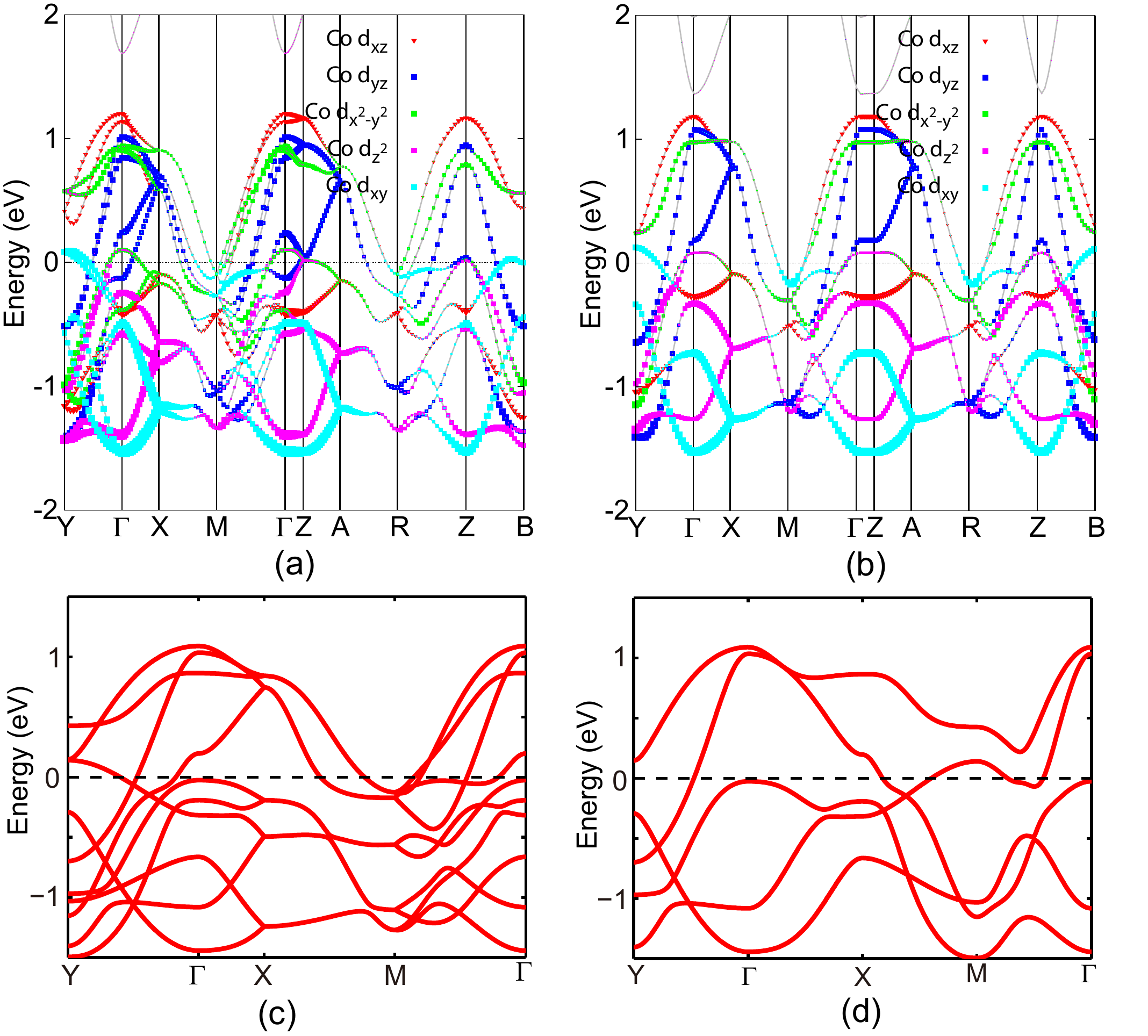}}
\caption{(color online) (a) and (b) are band structures of bulk and single layer BaCoSO. (c) The band structure of the effective model based on the five d-orbitals. (d) The unfold band structure of the effective model based on the five d-orbitals. \label{band} }
\end{figure}

We use  the density functional theory (DFT) under the generalized-gradient approximation (GGA) for the exchange correlation functional \cite{Kresse1993,Kresse1996,Kresse1996b,Perdew1996} to calculate the electronic band structures of the bulk BaCoSO and the  single layer BaCo(S,Se)O.   The optimized structural parameters of the single layers are given in the supplementary.  In Fig.\ref{band}, we plot the band structures of the bulk BaCoSO and the single layer BaCoSO. It is clear that the band dispersion along c-axis in the bulk material is small so that the main electronic structure is determined by a single CoAO layer.  The electronic structure also changes very little when S is replaced by Se (see the supplementary).  The $t_{2g}$  orbitals clearly dominate near Fermi energy in these materials.

We can fit the band structure to  a tight-binding Hamiltonian, $H_0$. For a single BaCoSO layer, the original unit cell has two Co atoms. However, similar to iron-based superconductors\cite{Lee2008}, because of the nonsymmorphic symmetry in the layer, the original two Co unit cell can be reduced into one Co unit cell by  a simple gauge transformation. Dividing the Co atoms in a single layer into two sublattices that are separated along the c-axis as shown in Fig.\ref{lattice}, we can choose a gauge which takes the Fermionic operators of d$_{xz/yz}$ orbitals to have alternating signs between two sublattices. Under this gauge, the band structure can be unfolded into an one Co unit cell.   The detailed fitting parameters are given in the supplementary for the unfolded $5\times 5$ matrix of $H_0$.  Fig.\ref{band} (c) and Fig.\ref{band}(d) show the fitted folded 10 d-orbital bands and  unfolded 5 d-orbital bands respectively.
We can further reduce $H_0$ by including only the three $t_{2g}$ orbitals. The three orbital model can capture the essential physics.  The detailed parameters and matrix elements for the 3 orbital model are also provided in the supplementary.

\begin{figure}
\centerline{\includegraphics[width=0.4\textwidth]{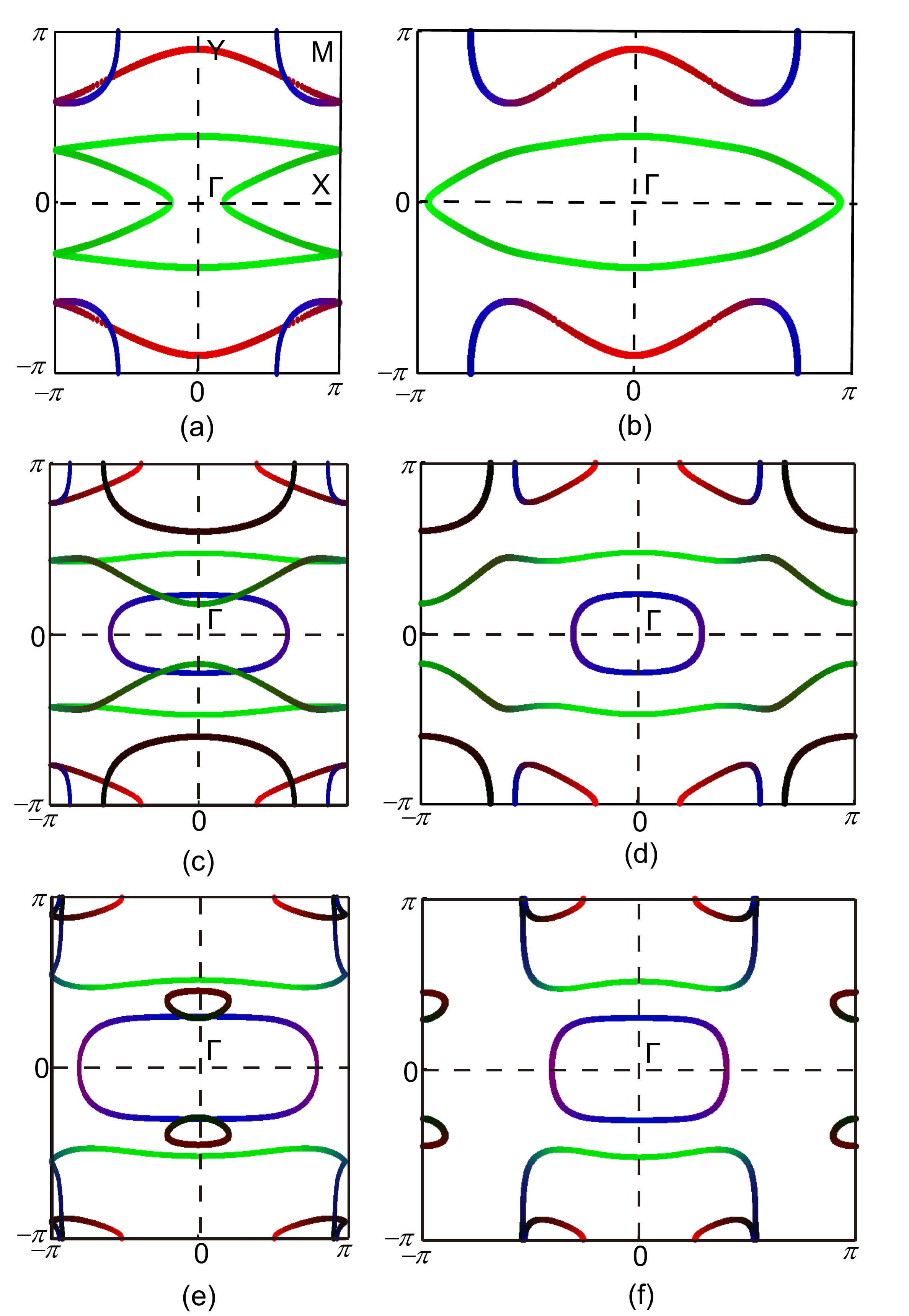}}
\caption{(color online)  Typical Fermi surfaces in both folded (left column) and unfolded BZ (right column)  from the effective model based on  all five d orbitals: (a,b), (c,d) and (e,f) are corresponding to  0.6 electron doping, half filling and 0.4 hole doping respectively.  The orbital contributions of the different FS sheets are shown color coded: d$_{xz}$(red), d$_{yz}$ (green) d$_{x^2-y^2}$ (bule) and e$_g$ orbitals (black).
\label{surface} }
\end{figure}
The tight binding parameters are very different from those derived for the previously proposed Co(S,Se)$_2$ layer\cite{Hu2016}.  In particular, the three t$_{2g}$ orbitals are organized and behave very differently. Because of the absence of C$_4$ symmetry, the band structure is quasi-one dimensional.  The d$_{yz}$ orbital is almost  decoupled from other two orbitals and produce a quasi-one dimensional Fermi surface that elongates along a-axis.  The d$_{xz}$ and d$_{x^2-y^2}$, however, are coupled with each other through oxygen along a-axis. In a simple picture, this coupling results in  binding and  antibinding bands. The mixing with the e$_g$ orbitals increases with hole doping. Combining the crystal energy splitting,   the topology of  Fermi surfaces produced by these two orbitals  highly depends on doping level.   Three typical Fermi surfaces are plotted in Fig.\ref{surface} in both folded BZ  and unfolded BZ. Fig.\ref{surface}(a,b), (c,d) and (e,f) are corresponding to 0.6 electron doping, half filling and 0.4 hole doping respectively in the effective model based on all five d orbitals.The doping levels refer to the offset from the half filling.

\begin{figure}
\centerline{\includegraphics[width=0.5\textwidth]{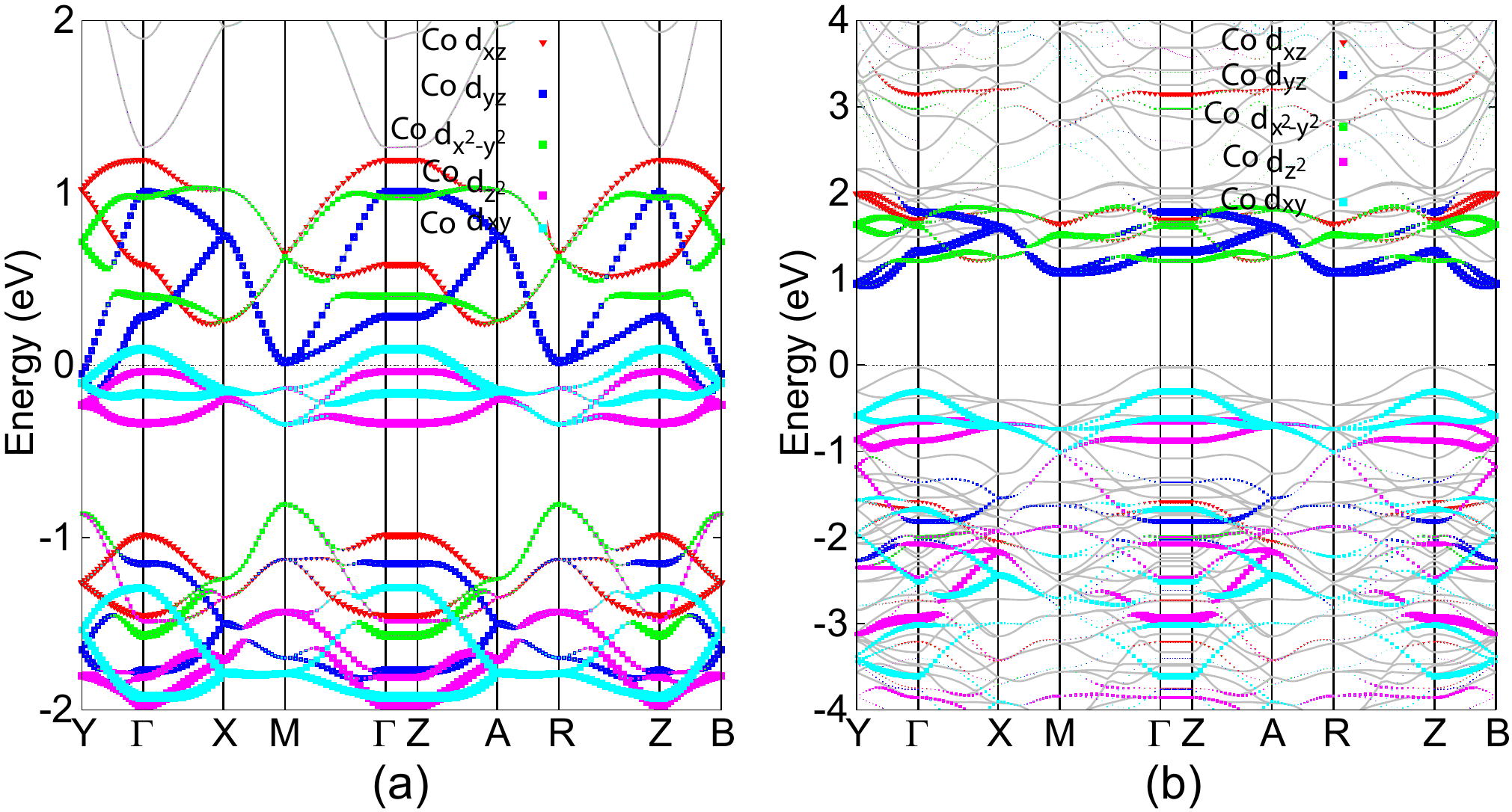}}
\caption{(color online) (a) The band structures of the single layer BaCoSO in the AFM state. (b) The band structures of the single layer BaCoSO under the GGA+U(U=3eV) calculations. The orbital characters of bands are represented by different colors.
\label{GGA} }
\end{figure}

The bulk BaCoSO has been measured experimentally to be an AFM insulator\cite{Salter2016,Valldor2015}.  The GGA+U calculations have also shown that with a reasonable U value, the material is in an AFM ordered state with a magnetic moment about 2.7$\mu_B$ per Co atom\cite{Valldor2015}. In Fig.\ref{GGA}, we confirm these results and plot the band structures without U and with U=3eV. Without U, the material  is   metallic  with a  magnetic moment about 1.7 $\mu_B$  per Co atom. The result indicates that the system is  a Mott insulator and the Co atoms are in the high spin state.   It is also clear that the magnetism here is stronger than that obtained in our previously proposed CoSe$_2$ layer\cite{Hu2016} because of the presence of oxygen atoms.

Including  the electron-electron correlation,  the full effective Hamiltonian for the material is given by $H= H_0+\sum_i H_i$ with
\begin{eqnarray}
H_{i}= \sum_{a} U n_{ia\uparrow}n_{ia\downarrow}+ \sum_{<ab>} (U'n_{ia }n_{ib}- J_H \vec S_{ia}\cdot\vec S_{ib}),
\end{eqnarray}
where the terms represent the Hubbard intra-orbital repulsion, inter-orbital repulsion and Hunds coupling respectively, and $a,b$ are orbital indices.  The value of U is expected to be around 3eV  in order to be consistent with  the measured magnetic moment, $J_H$ is around 0.9eV  and $U=U'+2J_H$.  Near half filling, as the magnetism is controlled by the AFM superexchange interactions, the effective Hamiltonian for effective magnetic interactions  can be written as
\begin{eqnarray}
H_{s}= \sum_{<ij>,\alpha} J^\alpha_{ab}  \vec S_{ia}\cdot\vec S_{jb},
\end{eqnarray}
where $<ij>$ is defined between two nearest neighbour(NN) sites and $\alpha=x,y$ which labels direction. $J^\alpha_{ab}$ are expected to be dominated by intra-orbital AFM couplings, namely $J^\alpha_{a=b}=J^{\alpha}_a$  are much larger than $J^\alpha_{a\neq b}$.  Combining $H_0$ with these effective interactions provides a minimum model to describe the system.  The value of $J^\alpha_{a}$ can be estimated from the energy of the AFM ordered state. For the single layer BaCoSO, we find that the total energy saving in the AFM ordered state is about 0.8eV per unit cell.  A rough estimation gives  the average magnetic exchange couplings for each orbital, which is about $J^x_a+J^y_a\sim 0.1$eV.

As the Fermi surfaces are dominated by the t$_{2g}$ orbitals which  are the  orbitals with the   AFM superexchange couplings, we expect that the superconductivity can be developed through the local AFM superexchange interactions if the AFM long range ordering can be suppressed by doping.  In particular, in this case,  it is easy to see that electron doping can reduce the effect of the Hunds coupling to suppress the AFM long range ordering quickly.   However, because of lacking of the C$_4$ rotational symmetry,   the simple argument based on  the Hu-Ding principle can not be applied here to extract the possible pairing symmetry.  In the following, we carry out a simple mean-field solution for  the effective t-J Hamiltonian, $H=H_0+H_s$. The similar calculation has been done for cuprates\cite{Kotliar1988} and iron-based superconductors\cite{Seo2008,Fang2011}.

\begin{figure}
\centerline{\includegraphics[width=0.45\textwidth]{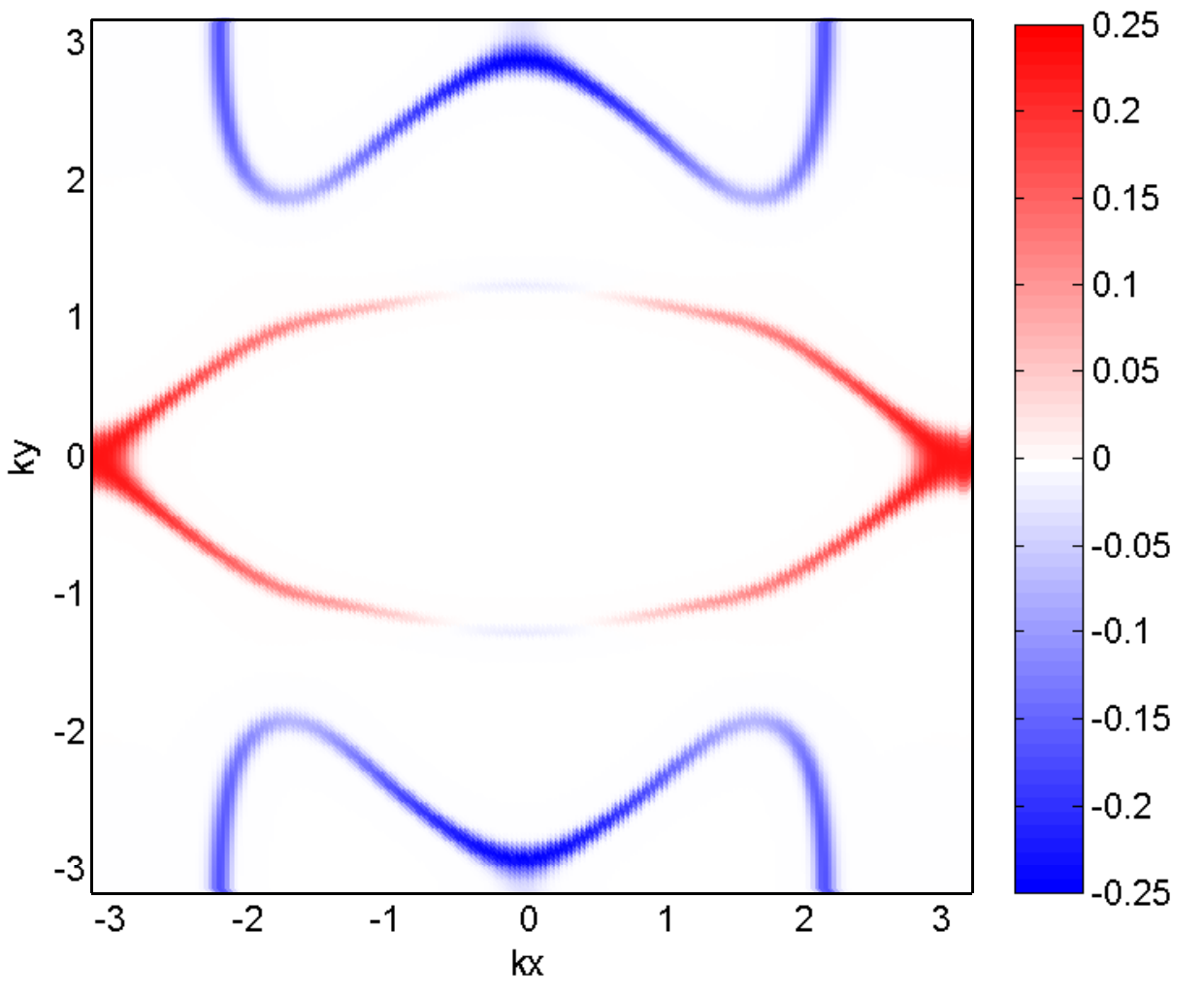}}
\caption{(color online) A typical mean field superconducting (SC) gaps in an arbitrary unit  and their sign distribution on the Fermi surfaces given in Fig.\ref{surface}(a,b).
\label{gap}}
\end{figure}

We focus on the electron doping region with Fermi surfaces as shown in Fig.\ref{surface}(a,b). As there is no $C_4$ rotational symmetry, the values of the AFM exchange couplings depend on orbitals and directions. Therefore, the parameter spaces are very large and it is hard to make quantitative predictions.  However, in all our calculations, we find that a d-wave like superconducting state is qualitatively  robust. In general,  for the Fermi surfaces given in Fig.\ref{surface}(a,b),   the sign of the superconducting order parameters  for intra-orbital $d_{xz}$ pairs and intra-orbital  d$_{yz} $ pairs are opposite and the sign of  the intra-orbital d$_{yz}$ pairs along two different directions are opposite which results in four gapless nodes on the Fermi surface with the d$_{yz}$ orbital characters.  A typical superconducting gap structure is given in Fig.\ref{gap} where we set the antiferromagnetic (AFM) coupling parameters as $J^x=0.8*J^y$ for all t$_{2g}$ orbitals.   More detailed calculations are summarized in the supplementary.  Such a d-wave like state is consistent with the Hu-Ding principle. As the pairing is dominated between two NN sites, the d-wave like structure can maximize the   absolute gap values on the Fermi surfaces.

 The absence of the C$_4$ rotational symmetry generally lowers T$_c$. The staggered structure of the Co atoms can also have negative effect on T$_c$ because lattice distortions can be easy to take place. However, as the energy scale  is close to those of iron-based superconductors, the maximum achievable $T_c$  should not be much lower than those of iron-based superconductors.  The main concern here is the intrinsic disorder that can be induced through doping.  With chemical doping,  both electron and hole doping can be introduced into BaCoAO  by substituting them with different atoms. However, the disorder induced by such a substitution may have much stronger effect on superconductivity than those in cuprates and iron-based superconductors because any substitution of these four atoms creates in-plane disorder. As there is  likely a sign change among superconducting order parameters in the superconducting state, the in-plane disorder can strongly suppress superconductivity.  Therefore, to achieve the  maximum T$_c$ in this family of materials, other doping techniques that have been widely applied to iron-based superconductors, such as  intercalating additional charge reservoir  layers\cite{Chen2015},  gate-tuning\cite{Ahn2006}, depositing alkaline atoms\cite{Yan2015}, and creating carrier transfer from a substrate\cite{Wang2012-fese}, may need to be implemented.

It is also interesting to note that we can  consider Ni-based oxychalcogenides, such as BaNiAO.  The hole doping to these Ni-based compounds, for example Ba$_{1-x}$K$_x$ NiAO,  is expected to achieve similar physics to the  electron doped Co-based oxychalcogenides. 

One popular theoretical method to explain superconductivity in correlated electron systems has been heavily relied on spin fluctuations generated through nesting between Fermi surfaces. Although this approach has been challenged in many iron-chalcogenides,  cuprates and iron-pnictides can be analyzed reasonably well by this method\cite{Hirschfeld2011}.
However, in these materials, there is no  nesting wavevector  on Fermi surfaces  that can be corresponding to  the AFM ordered wavevector due to a strong reorganization of the three t$_{2g}$ orbitals in the absence of the C$_4$ rotational symmetry.  Thus, experimental findings in the family can provide an ultimate judgement on the validity of this method.

In summary, we derive a low energy effective model  for the family of materials based on  the two dimensional CoAO layer.  The low energy physics is dominated by the three t$_{2g}$ orbitals  with strong AFM superexchange interactions.  We suggest that if  the AFM ordering can be suppressed upon doping or pressure,   a d-wave like superconducting state with reasonable high transition temperature can be developed. Confirming such a prediction can provide decisive evidence to  solve  the mechanism of unconventional high T$_c$ superconductivity.


{\it Acknowledgement:}  the work is supported by the Ministry of Science and Technology of China 973 program(Grant No. 2015CB921300), National Science Foundation of China (Grant No. NSFC-11334012), and   the Strategic Priority Research Program of  CAS (Grant No. XDB07000000).



%


\end{document}